%#!tex K3SYM.tex
%% Last Modified: Thu Jan 31 07:02:46 2008.

%\def\mydraft{jama}
%\def\mydraft{label}

\input lanlmac
\input amssym
\input epsf

%Macro for figure
\newcount\figno
\figno=0
\def\fig#1#2#3{
\par\begingroup\parindent=0pt\leftskip=1cm\rightskip=1cm\parindent=0pt
\baselineskip=13pt
\global\advance\figno by 1
\midinsert
\epsfxsize=#3
\centerline{\epsfbox{#2}}
\vskip 12pt
%\centerline{{\bf Fig. \the\figno:~~} #1}\par
{\bf Fig. \the\figno:~~} #1 \par
\endinsert\endgroup\par
}
\def\figlabel#1{\xdef#1{\the\figno}}
%%                              TABLEAUX.TEX
%%      This  macro file is for producing a ``Young Tableau'' which is
%%      an array of little squares sometimes used in mathematical physics.
%%      For instance, the command $\tableau{6 3 2}$ will produce a tableau
%%      with 6 squares in the top row, 3 in the next, and 2 in the last.
%%                                  OOOOOO
%%      This tableau will look like OOO    but made of squares instead of
%%           O's.
%%                                  OO
%%      Any number of rows may be present, each having a nonzero number of
%%      squares.
%%
%%      A tableau is math mode material, so use $ or $$ to enclose it.
%%
%%      The size and line-thickness of the little boxes are controlled by
%%  the
%%      dimension parameters --
%%              \tableauside=1.0ex              %(size)
%%              \tableaurule=0.4pt              %(line-thickness)
%%      Change them if you want.
%%
%%                                                      -- Doug Eardley
%%   9/19/8
%%
%%
\newdimen\tableauside\tableauside=1.0ex
\newdimen\tableaurule\tableaurule=0.4pt
\newdimen\tableaustep
\def\phantomhrule#1{\hbox{\vbox to0pt{\hrule height\tableaurule
width#1\vss}}}
\def\phantomvrule#1{\vbox{\hbox to0pt{\vrule width\tableaurule
height#1\hss}}}
\def\sqr{\vbox{%
  \phantomhrule\tableaustep

\hbox{\phantomvrule\tableaustep\kern\tableaustep\phantomvrule\tableaustep}%
  \hbox{\vbox{\phantomhrule\tableauside}\kern-\tableaurule}}}
\def\squares#1{\hbox{\count0=#1\noindent\loop\sqr
  \advance\count0 by-1 \ifnum\count0>0\repeat}}
\def\tableau#1{\vcenter{\offinterlineskip
  \tableaustep=\tableauside\advance\tableaustep by-\tableaurule
  \kern\normallineskip\hbox
    {\kern\normallineskip\vbox
      {\gettableau#1 0 }%
     \kern\normallineskip\kern\tableaurule}% 
  \kern\normallineskip\kern\tableaurule}}
\def\gettableau#1 {\ifnum#1=0\let\next=\null\else
  \squares{#1}\let\next=\gettableau\fi\next}

\tableauside=1.0ex
\tableaurule=0.4pt

%Macro

\def\th{\theta}

\def\o{\over}

\def\Si{\Sigma}

\def\riya{\rightarrow}

\def\bt{\beta}
\def\ga{\gamma}
\def\Ga{\Gamma}
\def\al{\alpha}

\def\rt#1{\sqrt{#1}}

\def\sitarel#1#2{\mathrel{\mathop{\kern0pt #1}\limits_{#2}}}

\def\cob{\delta}

%\newsec{References}
\lref\WittenKT{
  E.~Witten,
  ``Three-Dimensional Gravity Revisited,''
  arXiv:0706.3359 [hep-th].
  %%CITATION = ARXIV:0706.3359;%%
}
\lref\MaloneyUD{
  A.~Maloney and E.~Witten,
  ``Quantum Gravity Partition Functions in Three Dimensions,''
  arXiv:0712.0155 [hep-th].
  %%CITATION = ARXIV:0712.0155;%%
}
\lref\PapadodimasWM{
  K.~Papadodimas,
  ``S-duality and a large N phase transition in N = 4 SYM on K3 at strong coupling,''
  arXiv:hep-th/0510216.
  %%CITATION = HEP-TH/0510216;%%
}
\lref\MinahanVR{
  J.~A.~Minahan, D.~Nemeschansky, C.~Vafa and N.~P.~Warner,
  ``E-strings and N = 4 topological Yang-Mills theories,''
  Nucl.\ Phys.\  B {\bf 527}, 581 (1998)
  [arXiv:hep-th/9802168].
  %%CITATION = NUPHA,B527,581;%%
}
\lref\MaldacenaBW{
  J.~M.~Maldacena and A.~Strominger,
  ``AdS(3) black holes and a stringy exclusion principle,''
  JHEP {\bf 9812}, 005 (1998)
  [arXiv:hep-th/9804085].
  %%CITATION = JHEPA,9812,005;%%
}
\lref\LapanJX{
  J.~M.~Lapan, A.~Simons and A.~Strominger,
  ``Nearing the Horizon of a Heterotic String,''
  arXiv:0708.0016 [hep-th].
  %%CITATION = ARXIV:0708.0016;%%
}
\lref\ManschotZB{
  J.~Manschot,
  ``AdS$_3$ Partition Functions Reconstructed,''
  JHEP {\bf 0710}, 103 (2007)
  [arXiv:0707.1159 [hep-th]].
  %%CITATION = JHEPA,0710,103;%%
}
\lref\ManschotHA{
  J.~Manschot and G.~W.~Moore,
  ``A Modern Farey Tail,''
  arXiv:0712.0573 [hep-th].
  %%CITATION = ARXIV:0712.0573;%%
}
\lref\DijkgraafFQ{
  R.~Dijkgraaf, J.~M.~Maldacena, G.~W.~Moore and E.~P.~Verlinde,
  ``A black hole farey tail,''
  arXiv:hep-th/0005003.
  %%CITATION = HEP-TH/0005003;%%
}
\lref\VafaTF{
  C.~Vafa and E.~Witten,
  ``A Strong coupling test of S duality,''
  Nucl.\ Phys.\  B {\bf 431}, 3 (1994)
  [arXiv:hep-th/9408074].
  %%CITATION = NUPHA,B431,3;%%
}
\lref\KrausVU{
  P.~Kraus, F.~Larsen and A.~Shah,
  ``Fundamental Strings, Holography, and Nonlinear Superconformal Algebras,''
  JHEP {\bf 0711}, 028 (2007)
  [arXiv:0708.1001 [hep-th]].
  %%CITATION = JHEPA,0711,028;%%
}
\lref\SasakiVS{
  T.~Sasaki,
  ``Hecke operator and S-duality of N = 4 ADE gauge theory on K3,''
  JHEP {\bf 0307}, 024 (2003)
  [arXiv:hep-th/0303121].
  %%CITATION = JHEPA,0307,024;%%
}
\lref\LabastidaIJ{
  J.~M.~F.~Labastida and C.~Lozano,
  ``The Vafa-Witten theory for gauge group SU(N),''
  Adv.\ Theor.\ Math.\ Phys.\  {\bf 3}, 1201 (1999)
  [arXiv:hep-th/9903172].
  %%CITATION = 00203,3,1201;%%
}
\lref\YinAT{
  X.~Yin,
  ``On Non-handlebody Instantons in 3D Gravity,''
  arXiv:0711.2803 [hep-th].
  %%CITATION = ARXIV:0711.2803;%%
}
\lref\YinGV{
  X.~Yin,
  ``Partition Functions of Three-Dimensional Pure Gravity,''
  arXiv:0710.2129 [hep-th].
  %%CITATION = ARXIV:0710.2129;%%
}
\lref\GaiottoXH{
  D.~Gaiotto and X.~Yin,
  ``Genus Two Partition Functions of Extremal Conformal Field Theories,''
  JHEP {\bf 0708}, 029 (2007)
  [arXiv:0707.3437 [hep-th]].
  %%CITATION = JHEPA,0708,029;%%
}
\lref\GaberdielVE{
  M.~R.~Gaberdiel,
  ``Constraints on extremal self-dual CFTs,''
  JHEP {\bf 0711}, 087 (2007)
  [arXiv:0707.4073 [hep-th]].
  %%CITATION = JHEPA,0711,087;%%
}
\lref\HawkingDH{
  S.~W.~Hawking and D.~N.~Page,
  ``Thermodynamics Of Black Holes In Anti-De Sitter Space,''
  Commun.\ Math.\ Phys.\  {\bf 87}, 577 (1983).
  %%CITATION = CMPHA,87,577;%%
}
\lref\DabholkarYR{
  A.~Dabholkar,
  ``Exact counting of black hole microstates,''
  Phys.\ Rev.\ Lett.\  {\bf 94}, 241301 (2005)
  [arXiv:hep-th/0409148].
  %%CITATION = PRLTA,94,241301;%%
}
\lref\SenDP{
  A.~Sen,
  ``How does a fundamental string stretch its horizon?,''
  JHEP {\bf 0505}, 059 (2005)
  [arXiv:hep-th/0411255].
  %%CITATION = JHEPA,0505,059;%%
}
\lref\GaiottoZZ{
D. Gaiotto,
``Monster symmetry and Extremal CFTs,''
arXiv:0801.0988 [hep-th].
}
\lref\DijkgraafXW{
  R.~Dijkgraaf, G.~W.~Moore, E.~P.~Verlinde and H.~L.~Verlinde,
  ``Elliptic genera of symmetric products and second quantized strings,''
  Commun.\ Math.\ Phys.\  {\bf 185}, 197 (1997)
  [arXiv:hep-th/9608096].
  %%CITATION = CMPHA,185,197;%%
}
\lref\MooreET{
  G.~W.~Moore, N.~Nekrasov and S.~Shatashvili,
  ``D-particle bound states and generalized instantons,''
  Commun.\ Math.\ Phys.\  {\bf 209}, 77 (2000)
  [arXiv:hep-th/9803265].
  %%CITATION = CMPHA,209,77;%%
}

%%%%%%%%%%%%%%%%%%%%%%%%%%%%%%%%%%%%%%%%%%%%%%%%%%%%%%%%%%%%%%%%%
%                      Title Page                               %
%%%%%%%%%%%%%%%%%%%%%%%%%%%%%%%%%%%%%%%%%%%%%%%%%%%%%%%%%%%%%%%%%
\Title{             
}
{\vbox{
\centerline{${\cal N}=4$ SYM on K3 and the AdS$_3$/CFT$_2$ Correspondence}
}}

\vskip .2in

\centerline{Kazumi Okuyama}
\vskip5mm
\centerline{Department of Physics, Shinshu University}
\centerline{Matsumoto 390-8621, Japan}
\centerline{\tt kazumi@azusa.shinshu-u.ac.jp}
\vskip .2in

%\vskip 2cm
\vskip 3cm
\noindent

%%abstract
We study the Fareytail expansion of the topological partition function
of ${\cal N}=4$ $SU(N)$ super Yang-Mills theory on K3.
We argue that this expansion corresponds to
a sum over geometries in asymptotically AdS$_3$ spacetime, 
which is holographically dual to a large number of
coincident fundamental heterotic strings.

\Date{January 2008}
\vfill
\vfill

\newsec{Introduction}
The AdS/CFT correspondence is a powerful
way to study the quantum gravity with
a negative cosmological constant.
In particular, the AdS$_3$/CFT$_2$ correspondence
is interesting from
the viewpoint of quantum gravity since
three dimensional gravity has no propagating
degrees of freedom at the classical level,
hence the bulk theory might be simpler than the
higher dimensional cousins. 
Recently, Witten proposed a boundary CFT
which is dual to the pure gravity on $AdS_3$  \WittenKT\ (see also
\refs{\GaiottoXH\GaberdielVE\ManschotZB\YinGV\YinAT\MaloneyUD{--}\GaiottoZZ}).
It is found that the partition function of
boundary CFT has a nice interpretation
as the sum over geometries in the bulk.
However, there are some left-right
asymmetric contributions in the partition
functions which are difficult to
interpret semi-classically.
Moreover, the very existence of the pure gravity on $AdS_3$
as a quantum theory has not been established yet.
Therefore, it is desirable to study
$AdS_3$ gravity in the string theory setup.
The obvious problem is that the dual CFT is not known in general.
Even if the dual CFT is known, the partition function
is usually hard to compute. 

There are a few cases that we can study the AdS$_3$/CFT$_2$ correspondence
quantitatively. In \DijkgraafFQ, Type IIB theory on $AdS_3\times S^3\times
K3$ is studied by rewriting the partition function
of BPS states (elliptic genus) as a sum over geometries,
which is known as the Fareytail expansion.
The difficulty appeared in the pure gravity on $AdS_3$ is avoided
since  the elliptic genus depends only on the left movers.

In this paper, we study the partition function of ${\cal N}=4$ 
$SU(N)$ super Yang-Mills theory on K3. 
Via the string dualities, this is equal to the partition function
of BPS states of $N$ fundamental heterotic strings.
Using the technique 
in \refs{\DijkgraafFQ,\ManschotHA}, 
we show that this partition function
has an expansion as a sum over asymptotically $AdS_3$ geometries
and argue that they are dual to a large number of heterotic strings.
In section 2, we review the partition function of ${\cal N}=4$ 
SYM on K3 and its relation to the heterotic string.
In section 3, we write down the Fareytail expansion
of the partition function of ${\cal N}=4$ SYM on K3.
In section 4, we discuss some questions.

\newsec{${\cal N}=4$ SYM on K3 and Heterotic Strings: Review}
We first review the Vafa-Witten theory of topological ${\cal N}=4$
SYM \VafaTF\ and its relation to the BPS index of heterotic strings.

\subsec{${\cal N}=4$ SYM on K3}
In \VafaTF, it is shown that the 
topologically
twisted $SU(N)$ ${\cal N}=4$ SYM on K3 computes the generating
function of the Euler number of moduli space of $k$ instantons 
\eqn\Zinstsum{
Z_N(\tau)=\sum_{k=0}^\infty q^{k-N}\chi\Big({\cal M}_{N,k}(K3)\Big)
}
with $q=e^{2\pi i\tau}$.
The ${\cal N}=4$ $SU(N)$ SYM with $k$
instantons is realized by the following brane configuration
in Type IIA theory:
\eqn\Dfour{
N~ {\rm D}4~~ {\rm  on}~ ~K3\times {\Bbb R}_t\quad \oplus\quad k~ {\rm D}0~,
}
where ${\Bbb R}_t$ denotes the time direction.
In this brane picture, the shift $k\riya k-N$ of instanton number in
\Zinstsum\ is understood as the contribution of D0-brane charge from
the curvature of K3.

The partition function \Zinstsum\ is evaluated as follows.
Let us first consider the case of $U(1)$ gauge theory.
This is easily obtained by noting that the 
moduli space of $U(1)$ instantons is equal to the Hilbert scheme
of points on K3
\eqn\modHilb{
{\cal M}_{1,k}(K3)={\rm Hilb}^k(K3)~.
}
It is well-known that the 
cohomology of this space is given by
the 
Fock space of oscillators $\al_{-n}^A~(A=1\cdots24)$ at level $L_0=k$.
Note that $\al_{-1}^A$ corresponds to the generator of 
$H^0(K3)\oplus H^2(K3)\oplus H^4(K3)$
and the higher modes $\al_{-n}^A~(n>1)$ correspond to the
twisted sector of orbifold $(K3)^k/S_k$.
From this representation, one finds that 
the partition function of $U(1)$ theory
is given by the partition function of
24 free bosons
\eqn\Zone{
G(\tau)={1\over\eta(\tau)^{24}}~.
}
In the case of $SU(N)$ theory, the partition function is given
by an {\it almost} Hecke transform of the $U(1)$ partition function $G(\tau)$ 
\refs{\MinahanVR,\LabastidaIJ}
\eqn\ZNtau{
Z_N(\tau)={1\over N^2}\sum_{ad=N,b\in{\Bbb Z}_d}d\,G
\left({a\tau+b\over d}\right)~.
}
When $N=p$ is prime, this expression simplifies to
\eqn\Zprime{
Z_p(\tau)={1\o p^2}G(p\tau)+{1\o p}\sum_{b=0}^{p-1}G
\left({\tau+b\o p}\right)~.
}
As discussed in \refs{\VafaTF,\MinahanVR},
the structure of summation in \ZNtau\ can
be physically understood by adding mass term to
the adjoint scalar fields
and breaking the theory to $a$ factors of
${\cal N}=1$ $SU(d)$ pure Yang-Mills.
The summation over $b\in {\Bbb Z}_d$ comes from the
$d$ vacua of ${\cal N}=1$ $SU(d)$ theory.

Note that $Z_N(\tau)$ itself is not a modular form,
although $G(\tau)$ is a weight $-12$ modular form.
This is related to the fact that the Montonen-Olive S-duality maps
the $SU(N)$ theory to a theory with different gauge group $SU(N)/{\Bbb Z}_N$.
Therefore, $Z_N$ does not come back to itself under the action of
S-duality.

However, we can regard $Z_N$ as a member of
more general class of partition functions $Z_N^{(v)}$
with 't Hooft flux $v\in H^2(K3,{\Bbb Z}_N)$ turned on\foot{
One can introduce the theta series for the lattice $\Ga^{3,19}$
by summing over the 't Hooft fluxes.
This corresponds to considering $U(N)$ gauge theory
instead of $SU(N)$ gauge theory \VafaTF. 
},
and identify $Z_N=Z_N^{(v=0)}$.
The partition function with 't Hooft flux $v$ is given by
\SasakiVS
\eqn\fluxZ{
Z_N^{(v)}(\tau)={1\o N^2}\sum_{ad=N,b\in{\Bbb Z}_d}d\,G\left({a\tau+b\o d}\right)
\cob_{dv,0}e^{-\pi i{bv\cdot v\o aN}},
}
where $v\cdot v'=\int_{K3}v\wedge v'$ is the intersection number.
One can show that $Z_N^{(v)}$ transform as a vector-valued modular form
of weight $-12$ \SasakiVS
\eqn\ZNvga{
Z_N^{(v)}(\ga(\tau))=(c\tau+d)^{-12}\sum_{v'\in H^2(K3,{\Bbb Z}_N)}
M_{vv'}(\ga)Z_N^{(v')}(\tau)~.
} 
Throughout this paper we 
use the usual notation for $\ga\in SL(2,{\Bbb Z})$ and
its action on $\tau$
\eqn\gamat{
\ga=\left(\matrix{a&b\cr c&d}\right),\quad
\ga(\tau)={a\tau+b\o c\tau+d}~.
}
The modular matrix $M(\ga)$ for $S=\left(\matrix{0&1\cr -1&0}\right)$ and
$T=\left(\matrix{1&1\cr 0&1}\right)$ is given by
\eqn\MST{
M_{vv'}(S)={1\o N^{11}}e^{{2\pi i\o N}v\cdot v'},\quad
M_{vv'}(T)=\cob_{v,v'}e^{{\pi i\o N}v\cdot v}~.
}

It is instructive to explicitly write down 
the first few terms of 
$q$-expansion of partition functions \Zone, 
\ZNtau\ 
\eqn\Zexpand{\eqalign{
G&=q^{-1}+24+324q+3200q^2+25650q^3+\cdots,\cr
Z_2&={1\o4 }q^{-2}+30+3200q+176337q^2+5930496q^3+\cdots,\cr
Z_3&={1\o9}q^{-3}+{80\o3}+25650q+5930496q^2+639249408q^3+\cdots,\cr
Z_4&={1\o16}q^{-4}+{63\o2}+176256q+143184800q^2+42189811200q^3
+\cdots~.
}}
One immediately notices that 
$Z_N$ has a `gap' between $q^{-N}$ and $q^0$, {\it i.e.},
the coefficient of $q^n$ vanishes in the range $-N+1\leq n\leq -1$. 
This is true for
 general $N$: \foot{Curiously, the $q^0$ term of $Z_N$ is $24$ 
times the integral of matrix model obtained by the 
dimensional
reduction of $D=10$ super Yang-Mills to zero dimension \MooreET.}
\eqn\ZNconst{
Z_N={1\o N^2}q^{-N}+24\sum_{a|N}{1\o a^2}+{\cal O}(q)~.
}
The existence of `gap' is understood by counting the dimension of moduli space
\eqn\dimmoduli{
{\rm dim}\,{\cal M}_{N,k}(K3)=4N(k-N)+4~,
}
which becomes negative when $k<N$.
This implies that there is no contribution to $Z_N$
from the instantons with the instanton number $k<N$.
\subsec{Relation to Heterotic Strings}
By the duality chain, we can dualize the D4-D0 configuration in 
\Dfour\ to a configuration in heterotic string theory.
To see this, we first lift the IIA configuration \Dfour\ to 
the M-theory configuration:
\eqn\Mfive{
N~{\rm M}5~~ {\rm  on}~ ~K3\times {\Bbb R}_t\times S^1_{\rm M}\quad \oplus\quad k
~{\rm units~of~momentum~along}~S^1_{\rm M}~.
}
Here $S^1_{\rm M}$ denotes 
the M-theory circle in the eleventh direction.
In order to relate this configuration
to the topological ${\cal N}=4$ SYM, we perform a Wick rotation 
of the time direction ${\Bbb R}_t$ and compactify it to a thermal circle
$S_{\bt}^1$. Then the worldvolume of M5-brane becomes
$K3\times T^2$ where $T^2=S_{\bt}^1\times S_{\rm M}^1$.
More generally, we replace the two-dimensional part of
M5-brane worldvolume by a torus $\Si_\tau$ with an arbitrary modular
parameter $\tau$
\eqn\Etaurep{
{\Bbb R}_t\times S^1_{\rm M}\qquad
\longrightarrow\qquad{\rm Euclidean~torus}~~\Si_{\tau}~.
}
Using the relation between M5-brane compactified on a torus and
${\cal N}=4$ SYM, the moduli $\tau$ of torus $\Si_\tau$ is identified as
the coupling constant of ${\cal N}=4$ SYM
\eqn\tauYM{
\tau={\th\o2\pi}+i{4\pi\o g_{\rm YM}^2}~.
}

Finally, the relation between ${\cal N}=4$ SYM on K3 and the heterotic string
follows from the identification
of M5-brane wrapping around K3 and the fundamental heterotic string.
Therefore, the M5-brane configuration \Mfive\ is dual to
\eqn\hetframe{
N~{\rm heterotic~strings~on~}\Si_{\tau}\quad \oplus\quad
k~{\rm units~of~momentum~along}~S^1\subset \Si_{\tau}~.
}
In this heterotic string picture,
the partition function $Z_N$ is given by the index of BPS states
(Dabholkar-Harvey states)
in the ${\cal N}=(0,8)$ superconformal field theory of $N$
fundamental heterotic strings. 
This is computed by setting the right-moving SUSY part to the ground state
and summing over the left-moving bosonic side.
For the single string case, this summation gives $\eta(\tau)^{-24}$, as
expected from the result of $U(1)$ ${\cal N}=4$ SYM \Zone.
For $N>1$, the Hecke structure of $SU(N)$ SYM partition function
\ZNtau\ is interpreted in the heterotic picture
as the effect of multiple winding of genus one worldsheet around
the target space torus $\Si_\tau$ \refs{\MinahanVR,\DijkgraafXW}.

\newsec{Fareytail Expansion of ${\cal N}=4$ SYM on K3}
As discussed in \refs{\LapanJX,\KrausVU}, a large number of coincident
fundamental heterotic strings has
a near horizon geometry of the form
$AdS_3\times M$, hence it is expected to have a holographic dual
two-dimensional CFT. 
In the previous section, we saw that the partition function $Z_N$
of ${\cal N}=4$ SYM on K3 captures the BPS spectrum of 
$N$ fundamental heterotic strings.
Therefore, it seems natural to identify $Z_N$ as the BPS index of
string theory on the $AdS_3$ dual of heterotic strings.
Since we have Wick-rotated the time direction, the dual $AdS_3$ 
should be understood as the Euclidean $AdS_3$
and the torus $\Si_\tau$ is interpreted as the boundary
of $AdS_3$. The modular parameter $\tau$ should be fixed as 
a boundary condition
for the bulk metric.
Note that the large $N$ limit with $\tau$ fixed 
is different from the 't Hooft limit of ${\cal N}=4$ SYM,
which in turn implies that the AdS dual in question is not $AdS_5$ 
but $AdS_3$.

The Euclidean $AdS_3$ is topologically a solid torus.
There are many ways to fill inside the torus $\Si_\tau$
to make a solid torus. The bulk geometry is distinguished
by the homology cycle of $\Si_\tau$ which becomes
contractible. For instance,
the spatial circle is contractible for the thermal $AdS_3$
and the temporal circle is contractible for the BTZ black hole.

To see the relation of the partition function $Z_N$
to the bulk $AdS_3$ geometry\foot{The relation between 
the partition function $G(\tau)$ of
$U(1)$ theory and the black holes in ${\cal N}=4$ string 
theories is studied in \DabholkarYR.
The gravity dual of a single heterotic string is studied
in \SenDP.}, it is useful
to rewrite $Z_N$ as a Poincar\'{e} series.
A general procedure is developed in \refs{\DijkgraafFQ,\ManschotHA}
and dubbed Fareytail expansion.
The necessary ingredients are the modular matrix $M(\ga)$
in \ZNvga\ and the coefficient $c_v(n)$
of the polar part of 
$Z_N^{(v)}=\sum_{n}c_v(n)q^{n}$.
Applying the general formula in \ManschotHA\ to our case,
the Fareytail expansion of $Z_N$ reads\foot{The sum
over the coset $\Ga_\infty\backslash\Ga$
should be defined
as a limit \ManschotHA
$$\sum_{\Ga_\infty\backslash\Ga}\equiv
\lim_{K\riya\infty}\sum_{(\Ga_\infty\backslash\Ga)_K}=
\lim_{K\riya\infty}\sum_{|c|\leq K}\sum_{|d|\leq K,(c,d)=1}~.
$$
}
\eqn\farey{\eqalign{
Z_N(\tau)=12\sum_{a|N}{1\o a^2}&+{1\o2}\sum_{\ga\in \Ga_\infty\backslash\Ga}
(c\tau+d)^{12}\sum_{v\in H^2(K3,{\Bbb Z}_N)}M^{-1}(\ga)_{0v} \cr
&\times \sum_{n<0}c_v(n)\exp\left(2\pi in{a\tau+b\o c\tau+d}\right)
R\left({2\pi i|n|\o c(c\tau+d) }\right)~,
}}
where $\Ga_\infty=\left\{\left(\matrix{1&t\cr0&1}\right), 
t\in{\Bbb Z}\right\}$
is the parabolic subgroup of $\Ga=SL(2,{\Bbb Z})$,
and $R(x)$ is defined by
\eqn\Rxform{
R(x)={1\o(12)!}\int_0^x dt\,t^{12}e^{-t}~.
}

In the large $N$ limit, we expect that the expansion \farey\
can be interpreted as a sum over semi-classical geometries.
One can see that in the large $N$ limit
the summation over 't Hooft flux is dominated by the $v=0$
term, since the leading term of $Z_N^{(v\not=0)}$ is
$q^n$ with $n={\cal O}(N^0)$, while $Z_N^{(v=0)}$ starts with the term
${1\o N^2}q^{-N}$. 
Therefore, in the the large $N$ limit we can approximate $Z_N$ as 
\eqn\ZNasymp{ 
Z_N\sim {1\o2N^2}\sum_{\ga\in \Ga_\infty\backslash\Ga}
(c\tau+d)^{12}M^{-1}(\ga)_{00} 
\exp\left(-2\pi iN{a\tau+b\o c\tau+d}\right)
R\left({2\pi iN\o c(c\tau+d) }\right)~.
}
It seems natural to identify the exponential factor
in \ZNasymp\ as the holomorphic part of the classical action of
the $SL(2,{\Bbb Z})$ family of BTZ black holes \MaldacenaBW
\eqn\Sinst{
S=-4\pi N\,{\rm Im}\left({a\tau+b\o c\tau+d}\right)~.
}
Namely, the partition function $Z_N$ of
${\cal N}=4$ SYM on K3 admits a semi-classical expansion
of sum over geometries in the $AdS_3$ background, 
which is holographically dual
to heterotic strings.
As we move $\tau$ on the upper half plane, the 
dominant term in the sum \ZNasymp\ changes.
Since a large factor of $N$ is multiplied in the classical action
\Sinst, this change of dominant contribution
becomes a sharp phase transition
in the large $N$ limit.
This is interpreted as the Hawking-Page transition \HawkingDH\
in the bulk gravity side.
The phase diagram\foot{The phase diagram 
of ${\cal N}=4$ SYM on K3 was studied in \PapadodimasWM.
However, the motivation of \PapadodimasWM\ seems to be different
from ours.
} 
is the same as that of the pure
gravity on $AdS_3$ (see Fig.3b in \MaloneyUD).

\newsec{Discussion}
In this paper, we studied the Fareytail expansion
of the partition function of ${\cal N}=4$ SYM on K3
and interpreted it as a sum over geometries
dual to fundamental heterotic strings. 
It is observed in \PapadodimasWM\
that
the contribution of BTZ black hole
is reproduced by taking the saddle point of
instanton sum \Zinstsum.
To see this, recall that when the instanton number becomes large
the Euler number of
instanton moduli space scales as 
\eqn\chiCardy{
\chi\Big({\cal M}_{N,k}(K3)\Big)\sim e^{4\pi\rt{N(k-N)}}\qquad(k-N\gg1)~.
}
This essentially follows from the Cardy formula applied to the
$c=24N$ CFT. Then the partition function \Zinstsum\ is
approximated as
\eqn\ZNlargek{
Z_N\sim\sum_k e^{4\pi\rt{N(k-N)}}q^{k-N}~.
}
The saddle point $k=k_{0}$ of the above sum is given by
\eqn\saddlek{
k_{0}-N=-{N\o \tau^2}~,
}
and the value of the corresponding term turns out to be
\eqn\saddlevalue{
e^{4\pi\rt{N(k_{0}-N)}}q^{k_{0}-N}=e^{2\pi i{N\o\tau}}~.
}
One can see that the exponent is nothing but the classical action of
BTZ black hole. Therefore, it seems that the BTZ black hole corresponds to
a condensate of large number of instantons.
On the other hand, the zero-instanton term ${1\o N^2}q^{-N}$
corresponds to the thermal $AdS_3$.
It would be interesting to understand what happens
when adding $k_{0}$ units of momentum to the fundamental
heterotic string
and see what triggers the phase transition
in the heterotic string picture.
It would also be interesting to study the zeros of $Z_N(\tau)$
and see if the Hawking-Page transition is associated
with a condensation of Lee-Yang zeros \MaloneyUD.
Finally, it would be interesting to identify
the $(c_L,c_R)=(24N,12N)$ CFT of $N$ fundamental heterotic strings. 
\vskip5mm
\noindent
\centerline{\bf Acknowledgment} 
This work is supported in part by
MEXT Grant-in-Aid for Scientific Research \#19740135.

\listrefs
\bye